\documentclass[intlimits,twoside,a4paper]{article}

\usepackage[cp1251]{inputenc}

\usepackage{mathrsfs}
\usepackage[eqsecnum]{cmpj3}

\usepackage{bm}

\issue{2024}{27}{2}{23701}
\doinumber{10.5488/CMP.27.23701}

\title[Plasmonic effects in rod-like metal-dielectric nanoparticles]%
{Plasmonic effects in rod-like metal-dielectric nanoparticles%
}
\author[Ya. V. Karandas]{Ya. V. Karandas\orcid{0009-0008-7639-1091}\refaddr{label1,label2}\thanks{E-mail: \email{karzy@zp.edu.ua}.}}
\addresses{
\addr{label1} National University Zaporizhzhia Polytechnic, 64, Zhukovskogo Str., Zaporizhzhia 69063, Ukraine
\addr{label2} Zaporizhzhia Scientific Research Forensic Center of the MIA of Ukraine, 19-A Avaliani St., Zaporizhzhia 69068, Ukraine
}

\Keywords{polarizability, equivalent prolate spheroid, frequency dependencies, absorption and scattering cross-sections, rod-like nanoparticles}

\date{Received June 12, 2023, in final form December 2, 2023}

\begin{document}

\maketitle

\begin{abstract}
The optical properties of rod-like two-layer nanoparticles are studied using the notions of equivalent prolate spheroid. The calculations are presented for frequency dependencies for polarizability and the absorption and scattering cross-section of prolate spheroids, cylinders, and spherocylinders. The effect of the sizes, the shapes of the nanoparticle and the material of the core and the shell on the location of the maxima of the imaginary part of polarizability and absorption and scattering cross-sections is analysed. The recommendations regarding the shape and size ratio of the nanoparticles for obtaining the maximum value of radiation efficiency are formulated.
%
%
\printkeywords
%
\end{abstract}

\section{Introduction}

Recently, the study of plasmonic effects has become a booming field in nanophysics and nanotechnologies. In particular, the processes of  interaction between light and metallic nanoparticles of  different shapes are under an intensive study \cite{b1,b2,b3,b4,b5,b6}.

It is known that metallic nanoparticles significantly amplify the local electromagnetic fields near their surfaces at the surface plasmonic resonance (SPR) frequencies, which are determined by their morphology, geometry and dielectric permittivity of the environment. The plasmonic properties of the nanoparticles have potential applications in nanophotonics, biophotonics and biomedicine \cite{b7,b8,b9}, scanning near-field optical microscopy \cite{b10}, optical sensing and spectroscopic measurements \cite{b11,b12}.

\begin{figure}[!t]
\centerline{\includegraphics[width=0.50\textwidth]{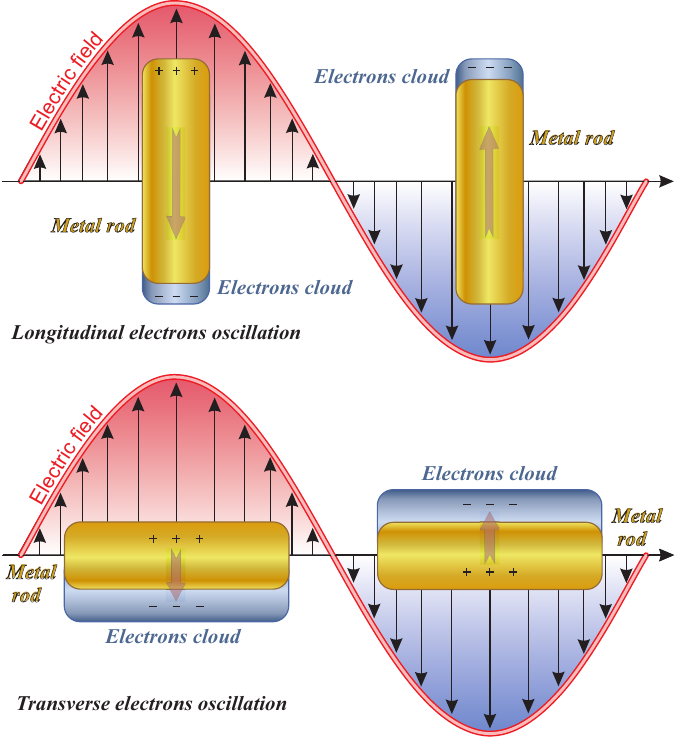}}
\caption{(Colour online) Diagrams of the excitation of longitudinal and transverse plasmonic resonances.} \label{fig-smp1}
\end{figure}

Despite the fact that biomedical applications of the plasmonic effects start from the use of the spherical nanoparticles, today in this field there is observed an increasing use of anisotropic nanoparticles, such as metallic spheroids and rods \cite{b13,b14,b15,b16,b17,b18}. Indeed, while spherical nanoparticles have one absorption peak, associated with localized surface plasmonic resonance (SPR), the localization of electrons along two dimensions in nanorods results in the emergence of two absorption peaks. One of these peaks is caused by the transverse SPR, and the other one –- by the longitudinal SPR (figure~\ref{fig-smp1}). As a rule, the longitudinal SPR band is located in the lower frequency range, which corresponds to the first or the second biological transparency window \cite{b18}, and it can be easily adjusted by adjusting an aspect ratio of the rod (the ratio of the radius to the length). As a result, the nanorods become prospective candidates for the use in biosensorics.

\begin{figure}[!t]
\centerline{\includegraphics[width=0.55\textwidth]{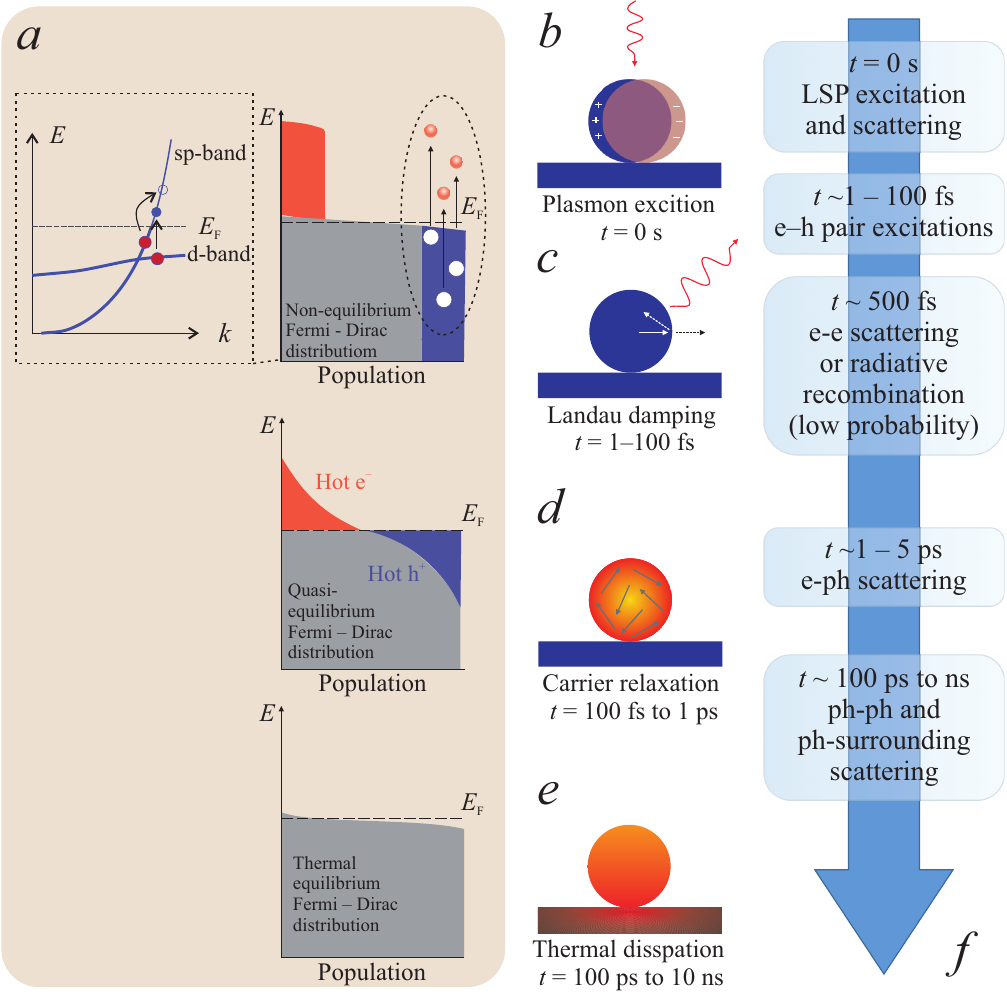}}
\caption{(Colour online) The mechanisms of electron-hole formation (a), relaxation processes, which determine the width of SPR line (b--d) and complete SPR relaxation schedule (f) on the example of spherical metallic nanoparticle of gold \cite{b5}.} 
\label{fig-smp2}
\end{figure}

One of the key spectral characteristics of SPR is its width, because it characterizes an enhancement of the optical response and local field inside and outside the particle. The width of SPR is determined by such dissipation mechanisms as volumetric and surface scattering and also radiation attenuation~\cite{b19,b20,b21}. The radiation attenuation is caused by the losses due to over-radiation of light and it becomes insignificant for  small metallic nanoparticles, which results in the narrowing of SPR spectrum to the limit set by the volumetric and surface scattering \cite{b20,b21}. They reflect Landau attenuation, that is decay of the charge collective oscillations into individual intrazone and interzone electron-hole pairs, and, hence, they are related to the intrinsic electronic properties of nanoparticles \cite{b19}. An interzone attenuation plays an important role when the localized SPR overlaps the interband transitions (figure~\ref{fig-smp2}), which results in a large widening of SPR, as in gold or copper nanospheres. When resonance is located away from the interband junction, as in silver nanospheres and gold nanorods for longitudinal SPR, it is much narrower~\cite{b22,b23}, which is of great interest for the applications. At the same time, the resonance width is determined by the electron relaxation rate, including the volumetric mechanisms (in particular, electron-electron and electron-photon scattering, and the scattering on defects) and the scattering on the surface, associated with the finite dimensions of the system \cite{b19}. The contribution of the last mechanism increases with a decrease of the particle size. Thus, the relaxation processes can be described in the form of a scheme, shown in figure~\ref{fig-smp2}, where electron population of Fermi level is shown in gray, hot electrons are represented by red regions above Fermi level $\varepsilon_{\rm{F}}$, and the distributions of the hot holes are represented by a blue region under $\varepsilon_{\rm{F}}$:

\begin{enumerate}
  \item the formation of electron-hole pairs in Au can be intrazone or interzone formation (figure~\ref{fig-smp2}a);
  \item after the excitation of SPR, the athermal distribution of electron-hole pairs relaxes during ${t}\sim1-100$~fs due to the emission of photons (radiologically) see figure~\ref{fig-smp2}{b} or due to Landau attenuation and formation of hot carriers (radiation-free) see figure~\ref{fig-smp2}c;
  \item hot carriers redistribute its energy due to electron-electron scattering on a time scale from $100$~fs  to 1~ps (figure~\ref{fig-smp2}d);
  \item then, the heat transfers inside the nanoparticle due to electron-photon scattering on a time scale of a few picoseconds and from the nanoparticle to the environment on the time scale of a few picoseconds to a few nanoseconds (figure~\ref{fig-smp2}e).
\end{enumerate}

\begin{figure}[!b]
	\centerline{\includegraphics[width=0.55\textwidth]{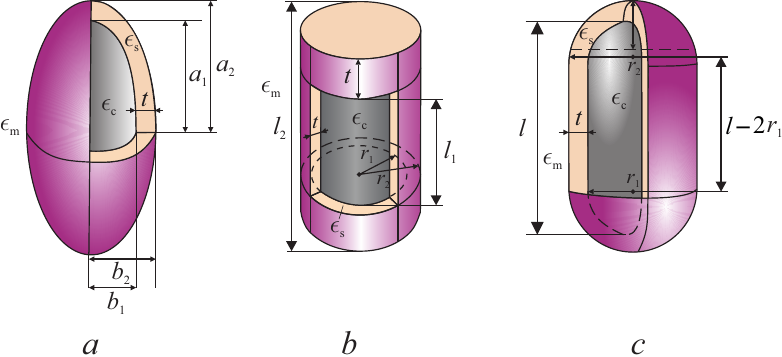}}
	\caption{(Colour online) Geometry of two-layer nanoparticles of different shapes which are under the study: a --- prolate spheroid; b --- cylinder; c --- spherocylinder.} 
	\label{fig-smp3}
\end{figure}

To date, the most fully studied model is the model of spherical nanoparticle \cite{b1}, but the model of spheroidal particle is more informative for the study of an influence of the nanoparticle shape on its optical properties. This is due to the fact that the above mentioned shape makes it possible to simulate the large class of particles of real shapes (from rod-like particles to disk-like particles), changing the ratio between the semi-axes of spheroid. At the same time, the change of the ratio between the semi-axes has a significant influence on the depolarization factors, which determine electromagnetic fields induced by the incident electromagnetic wave inside nanoparticle, and, correspondingly, on the electron scattering and optical absorption. A rigorous theory of attenuation in spheroidal metallic nanoparticles is possible only within the frameworks of using kinetic approach to the description of processes of electron scattering in the bulk and on the surface \cite{b24,b25,b26,b27,b28}. It is associated with the fact that the optical conductivity, which determines the half-width of the plasmonic resonances, becomes a tensor value in the case of spheroidal particles. Since the half-widths, in turn, determine the height of plasmonic peaks, then the shape of nanoparticles can have a significant effect on the frequency dependencies for light absorption and scattering cross-sections in spheroidal metallic nanoparticles \cite{b25}.

The model of prolate spheroid is used under the study of the optical properties of the rod-like nanoobjects. This is due to the fact that there is an analytical solution for the boundary problem of electrostatics for a spheroidal particle. The experimental results \cite{b29} demonstrate that the position of the longitudinal SPR for cylinders, spherocylinders and elongated spheroids depends on the shape of the nanoparticle. Moreover, it has been found that the model of the equivalent elongated spheroid gives a better approximation to the experimental results than the model of the elongated spheroid. This fact gives the authors of the work \cite{b30,b31} a right to state that the equalities of different axial moments of inertia for equivalent elongated spheroids and rod-like nanoparticles, which are under study, are indeed appropriate criteria for the comparison of rod-like nanostructures.

Coating the surface of the nanoparticle with an ultrathin dielectric shell, for example, with silicon dioxide, is an effective way to improve the compatibility and stability of the structure without a significant change of their optical properties \cite{b32,b33,b34}. This has a great practical significance for a successful use of plasmonic nanoparticles either in sensing or under the studies, connected with the surface such as surface-enhanced Raman light scattering (SERS). The presence of the silicon dioxide shell on the metallic particle allows them to have a well-controlled interface with the environment, in order to avoid the false widening effects \cite{b35}, which are observed, for example, for gold and silver nanospheres stabilized by surfactants \cite{b36}.

Thus, the aim of this article is to study the frequency dependencies for polarizability and the absorption and scattering cross-sections of metal-dielectric rod-like nanoparticles in the frameworks of the equivalent prolate spheroid approach.

\section{Basic relations}

Let us consider the rod-like nanoparticles with a metallic core which have the shape of elongated spheroid, cylinder with the finite length and spherocylinder covered with the dielectric shell with the thickness $t$ and permittivity $\epsilon_{\rm{s}}$, which are situated in the medium with the dielectric permittivity $\epsilon_{\rm{m}}$ (figure~\ref{fig-smp3}).

\begin{figure}[htb]
	\centerline{\includegraphics[width=0.40\textwidth]{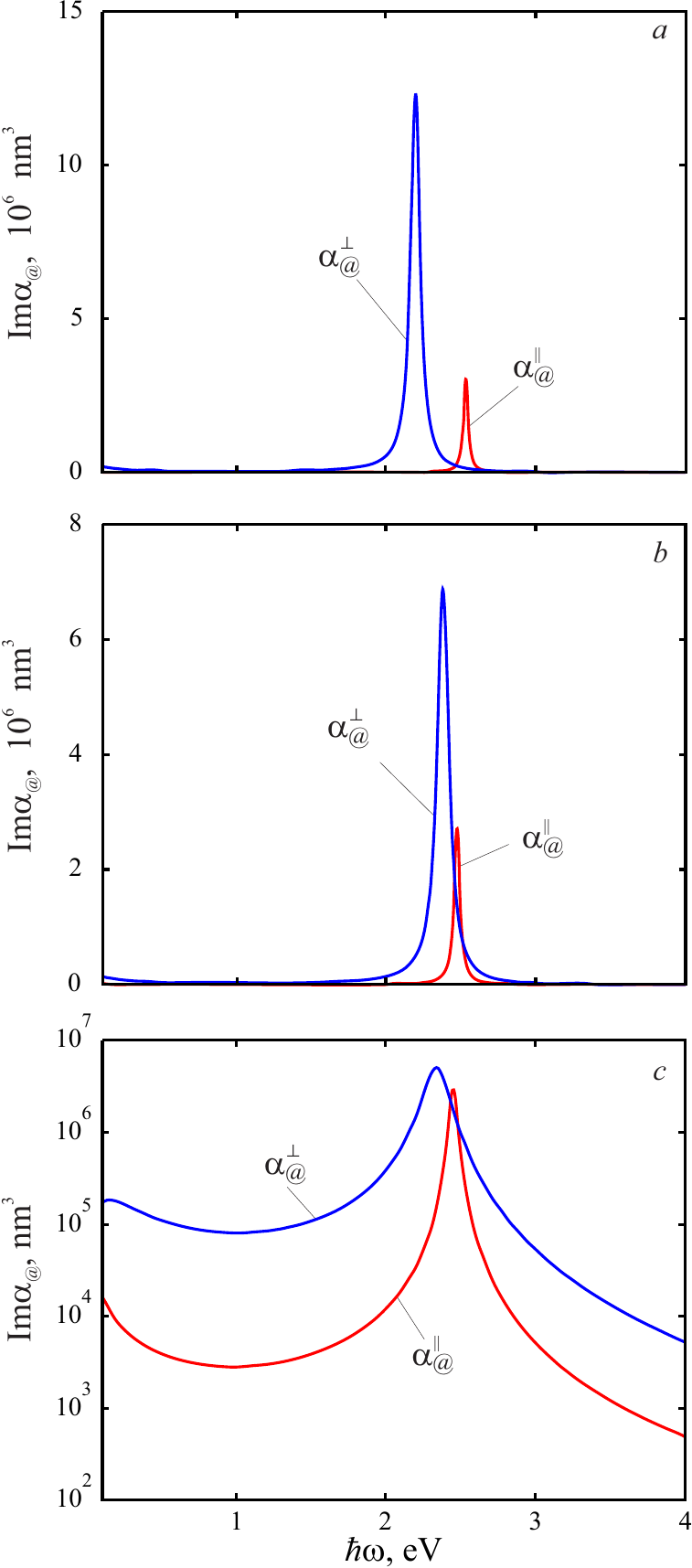}}
	\caption{(Colour online) The frequency dependencies for the imaginary parts of the longitudinal and transverse components of polarizability for nanoparticles $\rm{Au}@\rm{SiO}_2$ ($b_{t} = 20\,\,{\rm{nm}}$, ${a_l} = 150\,\,{\rm{nm}}$, $t = 5\,\,{\rm{nm}}$) of the shape of prolate spheroid (\emph{a}), cylinder (\emph{b}) and spherocylinder (\emph{c}).} 
	\label{fig-smp4}
\end{figure}

\begin{figure}[!t]
	\centerline{\includegraphics[width=0.40\textwidth]{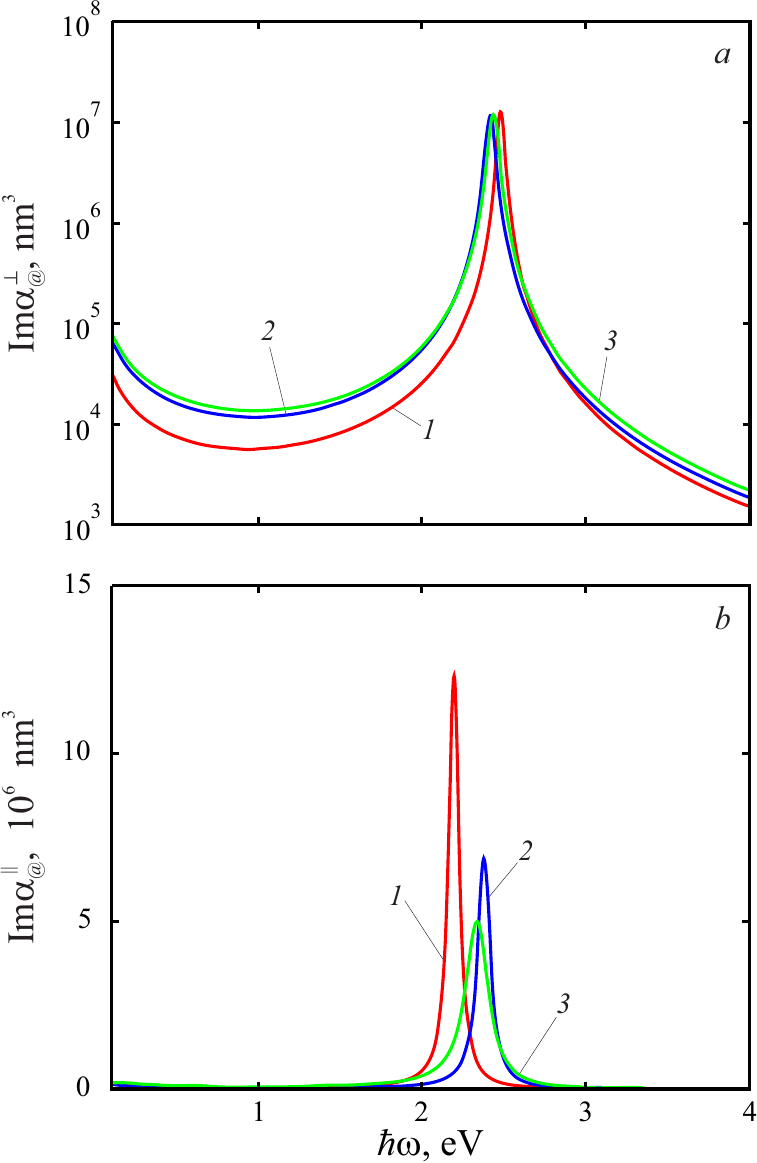}}
	\caption{(Colour online) The frequency dependencies for the transverse (\emph{a}) and longitudinal (\emph{b}) components of the imaginary parts of polarizability for the nanoparticles $\rm{Au}@\rm{SiO}_2$ with parameters $b_{t} = 20\,\,{\rm{nm}}$, ${a_l} = 150\,\,{\rm{nm}}$, $t = 5\,\,{\rm{nm}}$ of different geometrical shapes: 1 --- prolate spheroid; 2 --- cylinder; 3 --- spherocylinder.} 
	\label{fig-smp5}
\end{figure}

The starting point for the study are the relations for the diagonal components of polarizability of a two-layer prolate spheroidal particle \cite{b34,b37}

\begin{equation}
\label{q1}
  \alpha_@^{ \bot \left( \parallel  \right)} = V\frac{{\epsilon_@^{ \bot \left( \parallel  \right)}\, - \epsilon_{\rm{m}}}}{\epsilon_{\rm{m}} + \mathcal{L}_{ \bot \left( \parallel  \right)}^{\left( 2 \right)}\left( \epsilon_@^{ \bot \left( \parallel  \right)}\, - \epsilon_{\rm{m}} \right)},
\end{equation}
where $V$ is the volume of particle; $\epsilon_@^{ \bot \left( \parallel  \right)}$  is the diagonal components of dielectric tensor

\begin{equation}\label{q2}
\epsilon_@^{ \bot \left( \parallel  \right)} = \epsilon_{\rm{s}}^{ \bot \left( \parallel  \right)} + \frac{{\epsilon_{\rm{s}}^{ \bot \left( \parallel  \right)}{\beta _{\rm{c}}}\,\left( {\epsilon_{\rm{c}}^{ \bot \left( \parallel  \right)} - \epsilon_{\rm{s}}^{ \bot \left( \parallel  \right)}} \right)}}{{\epsilon_{\rm{s}}^{ \bot \left( \parallel  \right)} + \left( {\epsilon_{\rm{c}}^{ \bot \left( \parallel  \right)} - \epsilon_{\rm{s}}^{ \bot \left( \parallel  \right)}} \right)\left( \mathcal{L}_{ \bot \left( \parallel  \right)}^{\left( 1 \right)} - \beta _{\rm{c}}\mathcal{L}_{ \bot \left( \parallel  \right)}^{\left( 2 \right)} \right)}},
\end{equation}
and

\begin{equation}
\label{q3}
\mathcal{L}_\parallel ^{\left( {1,\,2} \right)} = \frac{{1 - e_p^{\left( {1,\,2} \right)\,2}}}{{2e_p^{\left( {1,\,2} \right)\,3}}}\left( {\ln \frac{{1 + e_p^{\left( {1,\,2} \right)}}}{{1 - e_p^{\left( {1,\,2} \right)}}} - 2e_p^{\left( {1,\,2} \right)}} \right), \,\,\,\,\,\,\, \mathcal{L}_ \bot ^{\left( {1,\,2} \right)} = \frac{1}{2}\left( {1 - \mathcal{L}_\parallel ^{\left( {1,\,2} \right)}} \right)
\end{equation}
are the depolarization factors of internal and external spheroid; $e_p^{\left( {1,\,2} \right)}$ are the corresponding eccentricities; $\epsilon_{\rm{s}}^ \bot  = \epsilon_{\rm{s}}^\parallel  = {\epsilon_{\rm{s}}} = {\mathop{\rm const}\nolimits} $ is the dielectric permittivity of the shell material; $\beta _{\rm{c}} = {V_{\rm{c}} \mathord{\left/
 {\vphantom {{{V_{\rm{c}}}} V}} \right. \kern-\nulldelimiterspace} V}$ is the bulk content of metallic fraction ($V_{\rm{c}}$ is the volume of metallic core), and the diagonal components of dielectric tensor of metallic core are determined by the following relations in the frameworks of Drude model

\begin{equation}
\label{q4}
\epsilon_{\rm{c}}^{ \bot \left( \parallel  \right)} =\epsilon^{\infty } - \frac{\omega_p^{2}}{{\omega \left( \omega  + {\mathop{\rm i}\nolimits} \gamma _{{\rm{eff}}}^{ \bot \left( \parallel  \right)}\right)}}.
\end{equation}
Here, $\epsilon^{\infty }$ is the component which describes the contribution of ion core; $\omega _{p} = {\left( {{e^2}{n_e}/{\epsilon_0}{m^*}} \right)^{{1 \mathord{\left/
 {\vphantom {1 2}} \right.
 \kern-\nulldelimiterspace} 2}}}$ is the plasma frequency, ${n_e}$ is the concentration of conductivity electrons, ${m^*}$ is the effective mass of electrons, ${\epsilon_0}$ is the electric constant; $\gamma _{{\rm{eff}}}^{ \bot \left( \parallel  \right)}$  are the transverse (longitudinal) effective relaxation rate, which is determined for the nanoscale objects as

\begin{equation}
\label{q5}
\gamma _{{\rm{eff}}}^{ \bot \left( \parallel  \right)} = {\gamma _{{\rm{bulk}}}} + \gamma _{\rm{s}}^{ \bot \left( \parallel  \right)} + \gamma _{{\rm{rad}}}^{ \bot \left( \parallel  \right)},
\end{equation}
where ${\gamma _{{\rm{bulk}}}}$ and $\gamma _{\rm{s}}^{ \bot \left( \parallel  \right)}$ are the bulk and surface relaxation rates, and $\gamma _{{\rm{rad}}}^{ \bot \left( \parallel  \right)}$ is the radiation damping rate.
\begin{figure}[!t]
	\centerline{\includegraphics[width=0.40\textwidth]{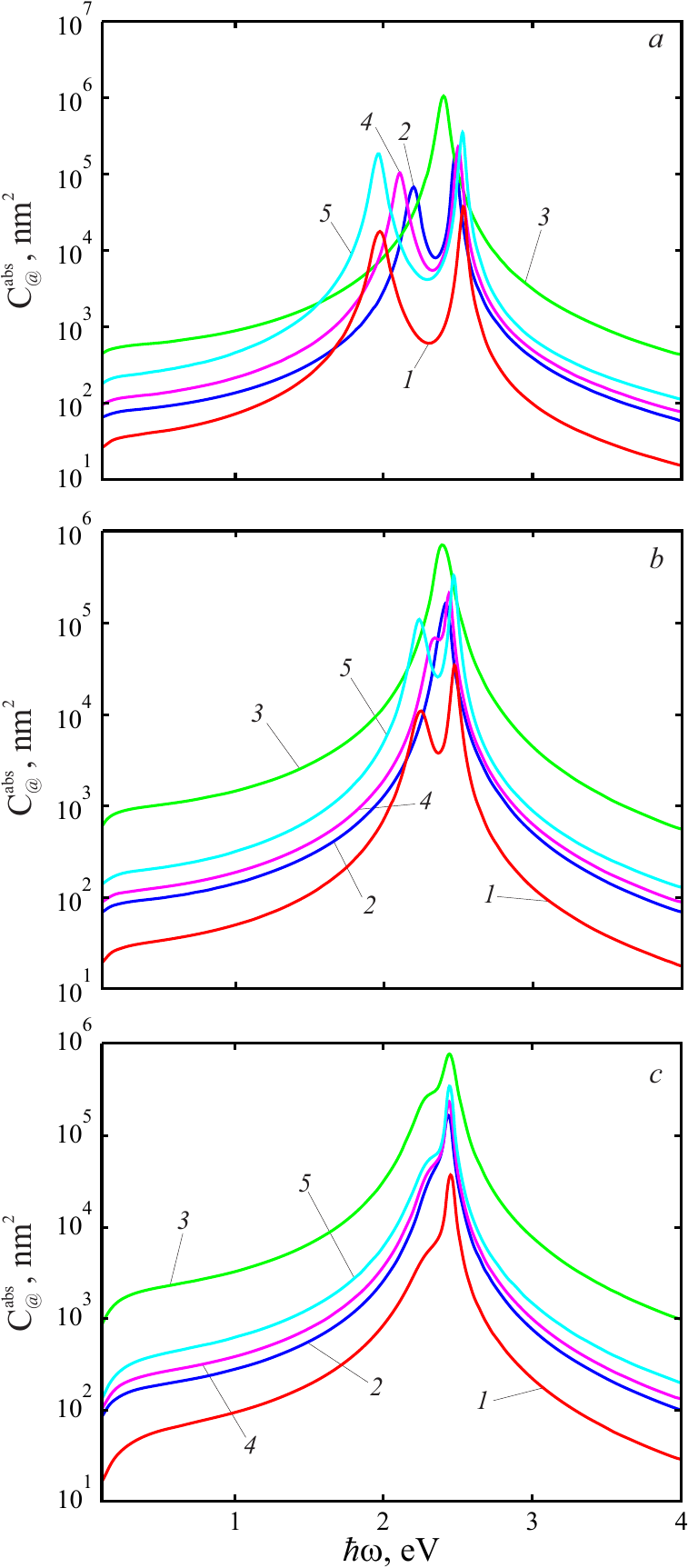}}
	\caption{(Colour online) The frequency dependencies for the absorption cross-section of the nanoparticles   of the shape of prolate spheroid (\emph{a}), cylinder (\emph{b}) and spherocylinder (\emph{c}) for $\rm{Ag}@\rm{SiO}_2$ with the different sizes: 1 --- $b_{t} = 10\,\,{\rm{nm}}$, ${a_l} = 150\,\,{\rm{nm}}$; 2 --- ${b_t} = 20\,\,{\rm{nm}}$, ${a_l} = 150\,\,{\rm{nm}}$; 3 --- ${b_t} = 50\,\,{\rm{nm}}$, ${a_l} = 150\,\,{\rm{nm}}$; \\4 --- ${b_t} = 20\,\,{\rm{nm}}$, ${a_l} = 200\,\,{\rm{nm}}$; 5 --- $b_{t} = 20\,\,{\rm{nm}}$, ${a_l} = 300\,\,{\rm{nm}}$.} 
	\label{fig-smp6}
\end{figure}

\begin{figure}[!t]
	\centerline{\includegraphics[width=0.40\textwidth]{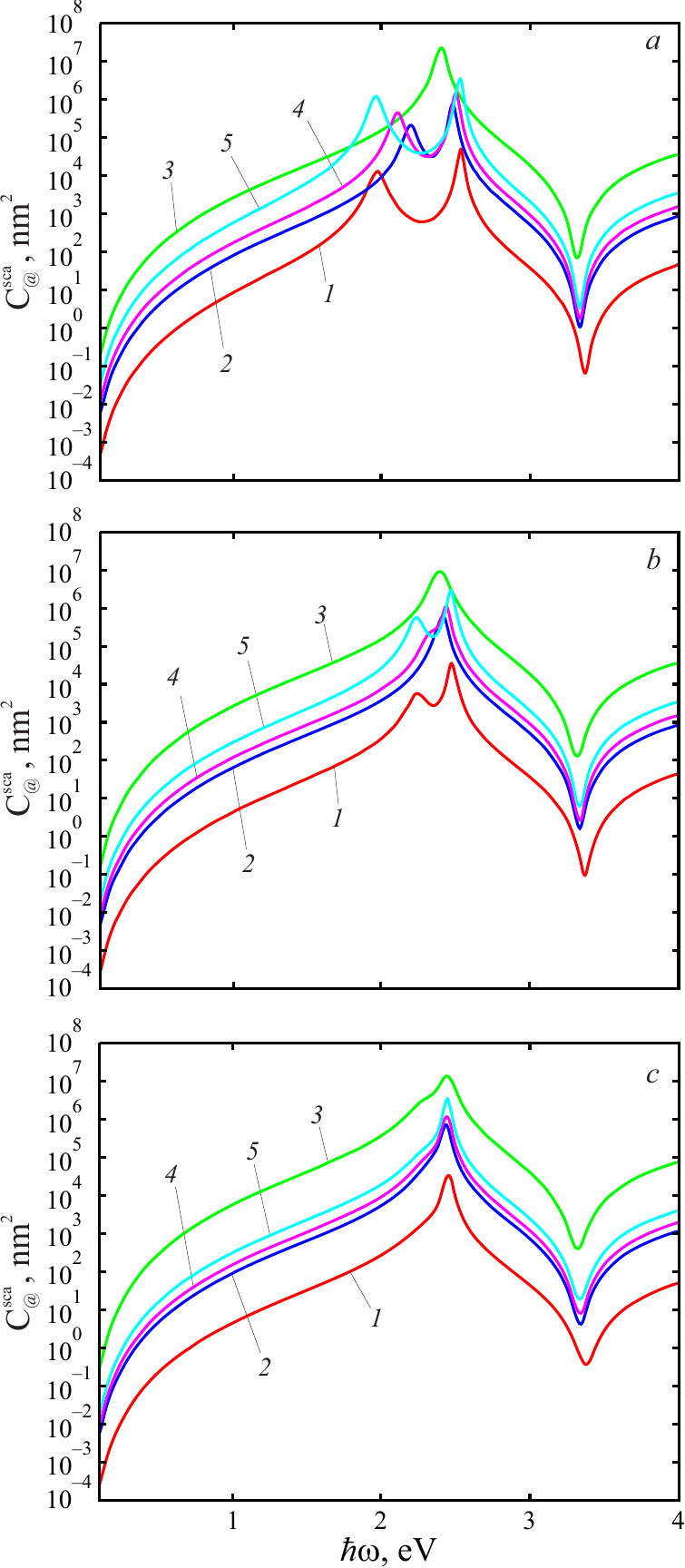}}
	\caption{(Colour online) The frequency dependencies for the scattering cross-sections of the nanoparticles $\rm{Ag}@\rm{SiO}_2$ of the shape of prolate spheroid (\emph{a}), cylinder (\emph{b}) and spherocylinder (\emph{c}) under the same values of parameters as in figure~\ref{fig-smp6}.} 
	\label{fig-smp7}
\end{figure}

\begin{figure}[!t]
	\centerline{\includegraphics[width=0.40\textwidth]{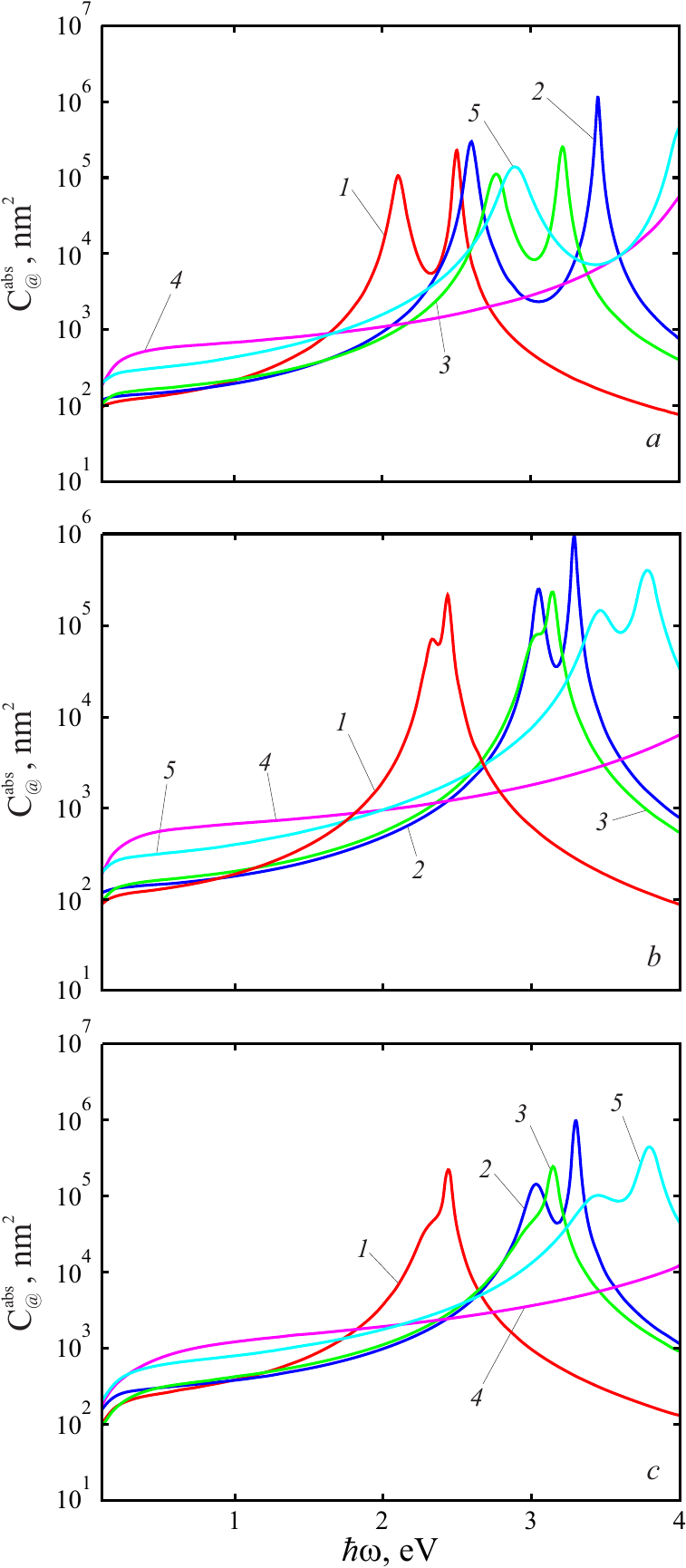}}
	\caption{(Colour online) The frequency dependencies for the absorption cross-sections of the nanoparticles $\rm{Me}@\rm{SiO}_2$ of the shape of prolate spheroid (\emph{a}), cylinder (\emph{b}) and spherocylinder (\emph{c}) under $b_{t} = 20\,\,{\rm{nm}}$, ${a_l} = 200\,\,{\rm{nm}}$, $t = 5\,\,{\rm{nm}}$ with the cores of different metals.} 
	\label{fig-smp8}
\end{figure}

\begin{figure}[!t]
	\centerline{\includegraphics[width=0.40\textwidth]{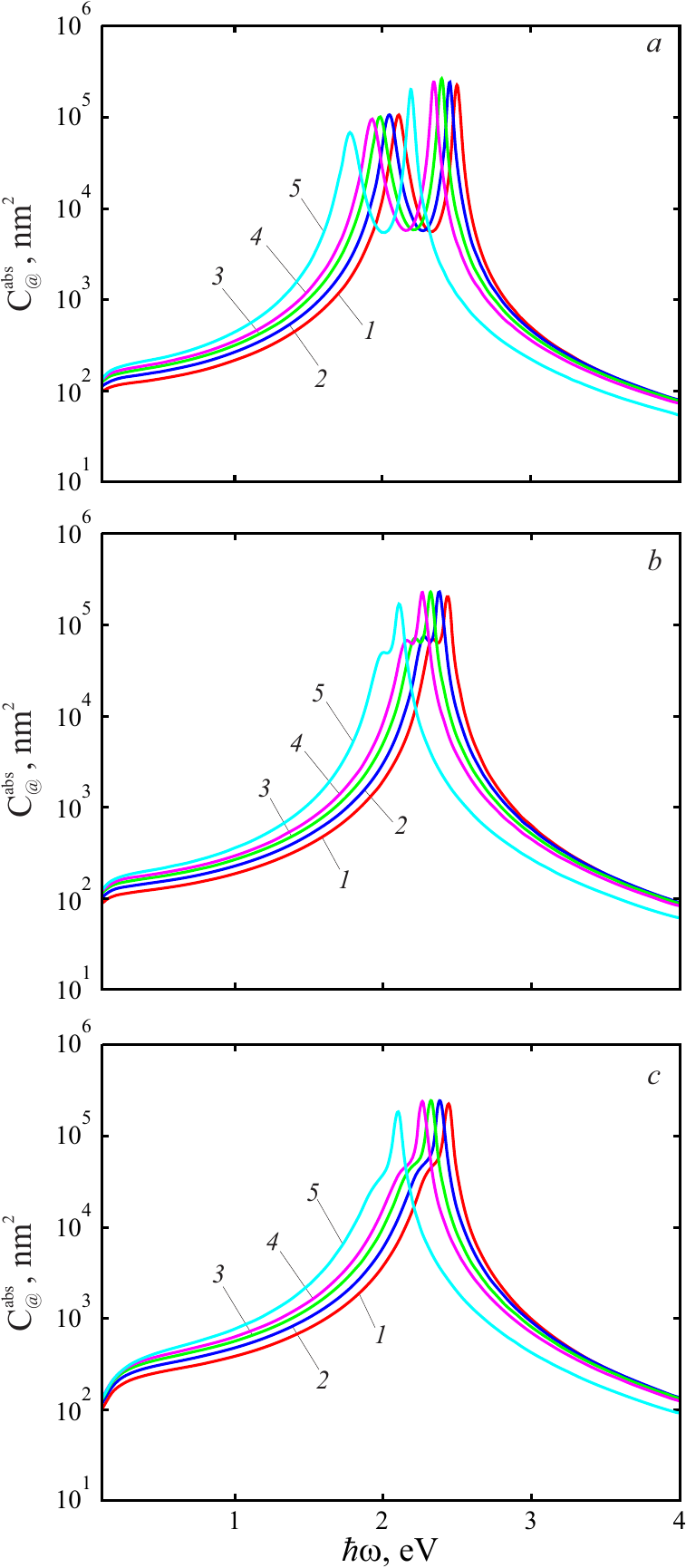}}
	\caption{(Colour online) The frequency dependencies for the absorption cross-sections of the nanoparticles ${\rm{Au}}@{\rm{D}}$ of the shape of prolate spheroid (\emph{a}), cylinder (\emph{b}) and spherocylinder (\emph{c}) under $b_{t} = 20\,\,{\rm{nm}}$, ${a_l} = 200\,\,{\rm{nm}}$, $t = 5\,\,{\rm{nm}}$ with the different shells: 1 --- ${\rm{Si}}{{\rm{O}}_2}$; 2 --- ${\rm{A}}{{\rm{l}}_2}{{\rm{O}}_3}$; 3 --- ${\rm{T}}{{\rm{a}}_2}{{\rm{O}}_5}$; 4 --- ${\rm{Ti}}{{\rm{O}}_2}$; 5 --- Si.} 
	\label{fig-smp9}
\end{figure}

\begin{figure}[!t]
	\centerline{\includegraphics[width=0.35\textwidth]{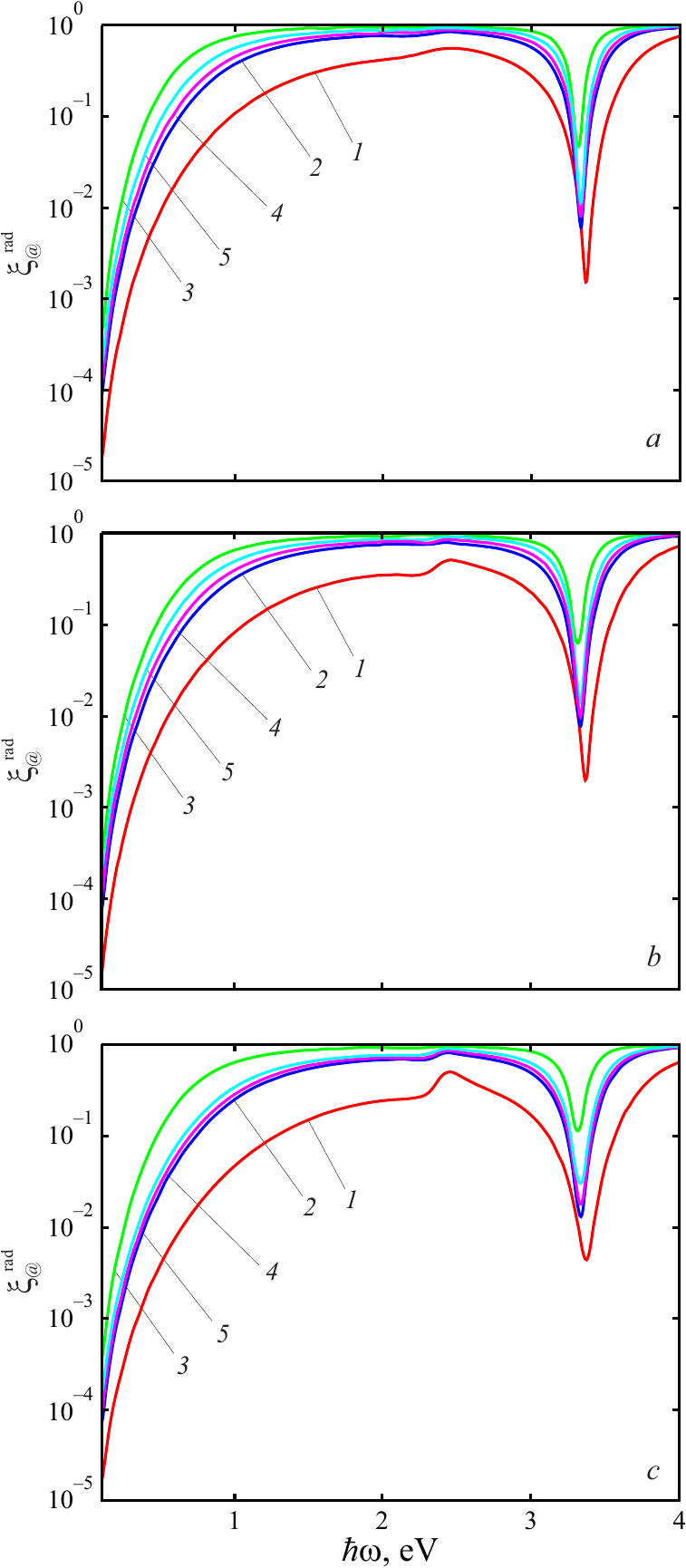}}
	\caption{(Colour online) The frequency dependencies for the radiation efficiency of the nanoparticles ${\rm{Au}}@{\rm{Si}}{{\rm{O}}_2}$ of the shape of prolate spheroid (\emph{a}), cylinder (\emph{b}) and spherocylinder (\emph{c}) under the same values of parameters as in figure~\ref{fig-smp6}.} 
	\label{fig-smp10}
\end{figure}

The surface relaxation rate and the radiation damping rate for prolate spheroid are determined by the relations \cite{b31}

\begin{equation}
\label{q6}
\gamma _{\rm{s}}^{ \bot \left( \parallel  \right)} = \frac{\mathcal{L}_{ \bot \left( \parallel  \right)}^{\left( 1 \right)}{\sigma _{ \bot \left( \parallel  \right)}}}{{{\epsilon_0}\left[ {{\epsilon_{\rm{m}}} + \mathcal{L}_{ \bot \left( \parallel  \right)}^{\left( 1 \right)}\left( 1 - \epsilon_{\rm{m}} \right)} \right]}};
\end{equation}

\begin{equation}
\label{q7}
\gamma _{\rm{rad}}^{ \bot \left( \parallel  \right)} = \frac{{2{V}}}{{9\piup\epsilon_{0}}}{\left( {\frac{\omega_{p}}{c}} \right)^3}\frac{\mathcal{L}_{ \bot \left( \parallel  \right)}^{\left( 1 \right)}{\sigma _{ \bot \left( \parallel  \right)}}}{{\sqrt {{\epsilon_{\rm{m}}}\left[ {{\epsilon^\infty } + \left( {\frac{1}{\mathcal{L}_{ \bot \left( \parallel  \right)}^{\left( 1 \right)}} - 1} \right){\epsilon_{\rm{m}}}} \right]} }}\,,
\end{equation}
where $c$ is the speed of light.

Formulae (\ref{q6}) and (\ref{q7}) possess  diagonal components of the conductivity tensor for a prolate spheroid, which can be determined with the help of the relations obtained in works \cite{b26,b27,b28}:

\begin{equation}
\label{q8}
\sigma_{\parallel } = \frac{{9{n_e}{e^2}}}{{2{m^*}\omega }}{\left( {\frac{\omega }{{{\nu _{s,\, \bot }}}}} \right)^2}\frac{1}{{e_p^{\left( 1 \right)\,3}}}\int\limits_{\frac{\omega }{{{\nu _{s,\, \bot }}}}}^{\frac{\omega }{{{\nu _{s,\,\parallel }}}}} {\frac{{\rd x}}{{{x^4}}}{{\left[ {1 - {{\left( {\frac{\omega }{{{\nu _{s,\, \bot }}x}}} \right)}^2}} \right]}^{\frac{1}{2}}}} \left\{ 1 - \frac{2}{x}\sin x + \frac{2}{{{x^2}}}\left( 1 - \cos x \right) \right\},
\end{equation}

\begin{equation}
\label{q9}
{\sigma _ \bot } = \frac{{9{n_e}{e^2}}}{{4{m^*}\omega }}{\left( {\frac{\omega }{{{\nu _{s,\, \bot }}}}} \right)^2}\frac{{e_p^{\left( 1 \right)\,2} - 1}}{{e_p^{\left( 1 \right)\,3}}}\int\limits_{\frac{\omega}{{{\nu _{s,\, \bot }}}}}^{\frac{\omega }{{{\nu _{s,\,\parallel }}}}} {\frac{{\rd x}}{{{x^4}}}\frac{{1 - {{\left( {\frac{\omega }{{{\nu _{s,\,\parallel }}x}}} \right)}^2}}}{{{{\left[ {1 - {{\left( {\frac{\omega }{{{\nu _{s,\, \bot }}x}}} \right)}^2}} \right]}^{\frac{1}{2}}}}}} \left\{ {1 - \frac{2}{x}\sin x + \frac{2}{{{x^2}}}\left( {1 - \cos x} \right)} \right\},
\end{equation}
where ${\nu _{s,\, \bot }}$ and ${\nu _{s,\,\parallel }}$ are the frequencies of individual oscillations of electrons along the axes of spheroid.

Let us determine, in the frameworks of the equivalent prolate spheroid approach, the effective aspect ratios for two-layer elongated spheroid, which is equivalent to cylinder of the finite length and to spherocylinder. For this purpose, let us write down the expressions for the axial moments of inertia of these objects:

\begin{equation*}
\label{1111}
I_{x,\,i}^{{\rm{ell}}} = \frac{{m_{{\rm{ell}}}^{\left( i \right)}}}{5}\left( {a_i^2 + b_i^2} \right);\,\,\,\,
I_{z,\,i}^{{\rm{ell}}} = \frac{{2m_{{\rm{ell}}}^{\left( i \right)}}}{5}b_i^2;
\end{equation*}
\begin{equation}
\label{q10}
I_{x,\,i}^{{\rm{cyl}}} = \frac{{m_{{\rm{cyl}}}^{\left( i \right)}}}{{12}}\left( {3r_i^2 + l_i^2} \right);\,\,\,\,
I_{z,\,i}^{{\rm{cyl}}} = \frac{{m_{{\rm{cyl}}}^{\left( i \right)}}}{2}r_i^2;
\end{equation}
\begin{equation*}
\label{1111}
I_{x,\,i}^{{\rm{sphcyl}}} = \piup r_i^5{\mu _i}\left\{ {\frac{{{\delta _i}}}{6}\left( {3 + \delta _i^2} \right) + \frac{4}{3}\left[ {\frac{{83}}{{320}} + {{\left( {{\delta _i} + \frac{3}{8}} \right)}^2}} \right]} \right\};\,\,\,\,
I_{z,\,i}^{{\rm{sphcyl}}} = \piup r_i^5{\mu _i}\left( \delta _{i} + \frac{8}{{15}} \right);
\end{equation*}
\begin{equation*}
\label{1111}
{\delta _i} = 1 - {\varrho _i},\quad
i = 1,\,\,2,
\end{equation*}
where indices 1 and 2 are associated with metallic core and the entire particle, correspondingly; $a_{2} =a_{1} + t$, ${b_2} = {b_1} + t$ are the major and minor semi-axes of spheroid; $l_{2} =l_{1} + 2t$ is the length of cylinder, $r_{2} =r_{1} + t$ is the radius of cylinder (spherocylinder); $m_{\rm{ell}}^{\left( i \right)}$ and $m_{{\rm{cyl}}}^{\left( i \right)}$ are the masses of ellipsoid and cylinder; ${\mu _i}$ is the density of spherocylinder.

Using the relations

\begin{equation}
\label{q11}
\frac{{I_{x,\,i}^{{\rm{ell}}}}}{{I_{z,\,i}^{{\rm{ell}}}}} = \frac{{I_{x,\,i}^{{\rm{cyl}}}}}{{I_{z,\,i}^{{\rm{cyl}}}}};
\,\,\,\,
\frac{{I_{x,\,i}^{{\rm{ell}}}}}{{I_{z,\,i}^{{\rm{ell}}}}} = \frac{{I_{x,\,i}^{{\rm{sphcyl}}}}}{{I_{z,\,i}^{{\rm{sphcyl}}}}},
\end{equation}
and introducing the aspect ratios for prolate spheroid, cylinder and spherocylinder

\begin{equation}
\label{q12}
{\varrho ^{\left( i \right)}} = \frac{{{b_i}}}{{{a_i}}};
\,\,\,\,
{\varrho ^{\left( i \right)}} = \frac{{2{r_i}}}{{{l_i}}};
\,\,\,\,
{\varrho ^{\left( i \right)}} = \frac{{2{r_i}}}{l + 2r_{i}}.
\end{equation}

Let us obtain the effective aspect ratios for spheroids which are equivalent to cylinder and spherocylinder:

\begin{equation}
\label{q13}
\varrho _{{\rm{eff}}}^{\left( i \right)} = \frac{{\sqrt 3 }}{2}{\varrho ^{\left( i \right)}};
\,\,\,\,
\varrho _{{\rm{eff}}}^{\left( i \right)} = {\left( {1 + \frac{{4{\delta _i}}}{3}\frac{{\delta _i^2 + 2{\delta _i} + \frac{3}{4}}}{{{\delta _i} + \frac{8}{{15}}}}} \right)^{ - \frac{1}{2}}}.
\end{equation}

Calculating integrals (\ref{q8}) and (\ref{q9}) under the condition of neglecting the oscillating terms compared to the unit in braces, we obtain the expression for the diagonal components of conductivity tensor

\begin{equation}
\label{q14}
{\sigma _{ \bot \left( \parallel  \right)}} = \frac{9}{{16}}{\epsilon_0}{\left( {\frac{{{\omega _p}}}{\omega }} \right)^2}{\nu _{s,\, \bot }}{\mathscr{F}_{ \bot \left( \parallel  \right)}}\left( {\varrho _{{\rm{eff}}}^{\left( 1 \right)}} \right),
\end{equation}
where

%
\begin{eqnarray}
	\label{q15}
	{\mathscr{F}_\bot}\left( {\varrho_{\rm{eff}}^{( 1 )}} \right) &=& \left( 1 - \left(\varrho_{\rm eff}^{(1)}\right)^2 \right)^{ - \frac{3}{2}}\left\{ \varrho _{\rm{eff}}^{\left( 1 \right)}\left( {\frac{3}{2} -\left(\varrho_{\rm eff}^{(1)}\right)^2} \right)\sqrt {1 - {\varrho_{\rm{eff}}^{\left( 1 \right)\,2}}}  \right.
		\nonumber\\
		&+& \left. 2 \left( {\frac{3}{4} - \left(\varrho_{\rm eff}^{(1)}\right)^2} \right)\ln \left( \frac{\piup }{2} - \arcsin \varrho _{\rm{eff}}^{( 1 )} \right) \right\};
\end{eqnarray}

\begin{equation}
\label{q16}
{\mathscr{F}_\parallel }\left( {\varrho _{\rm{eff}}^{\left( 1 \right)}} \right) = {\left( {1 - \left(\varrho_{\rm eff}^{(1)}\right)^2} \right)^{ - \frac{3}{2}}}\left\{ {\frac{\piup }{2} - \arcsin \varrho _{\rm{eff}}^{\left( 1 \right)} + \varrho _{\rm{eff}}^{\left( 1 \right)}\left( {1 - 2\left(\varrho_{\rm eff}^{(1)}\right)^2} \right)\sqrt {1 - \left(\varrho_{\rm eff}^{(1)}\right)^2} } \right\}.
\end{equation}

Substituting formula (\ref{q14}) into the relations (\ref{q6}) and (\ref{q7}), we obtain the general expressions for the surface relaxation rate and radiation damping rate

\begin{equation}
\label{q17}
\gamma _{\rm{s}}^{ \bot \left( \parallel  \right)} = \frac{9}{{16}}\frac{\mathcal{L}_{ \bot \left( \parallel  \right)}^{\left( 1 \right)}}{{{\epsilon_{\rm{m}}} + \mathcal{L}_{ \bot \left( \parallel  \right)}^{\left( 1 \right)}\left( {1 - {\epsilon_{\rm{m}}}} \right)}}{\nu _{s,\, \bot }}{\left( {\frac{{{\omega _p}}}{\omega }} \right)^2}{\mathscr{F}_{ \bot \left( \parallel  \right)}}\left( {\varrho _{{\rm{eff}}}^{\left( 1 \right)}} \right);
\end{equation}

\begin{equation}
\label{q18}
\gamma _{{\rm{rad}}}^{ \bot \left( \parallel  \right)} = \frac{V}{{8\piup }}\frac{\mathcal{L}_{ \bot \left( \parallel  \right)}^{\left( 1 \right)}}{{\sqrt {{\epsilon_{\rm{m}}}\left[ {{\epsilon^\infty } + \left( \frac{1}{\mathcal{L}_{ \bot \left( \parallel  \right)}^{\left( 1 \right)}} - 1 \right){\epsilon_{\rm{m}}}} \right]} }}{\nu _{s,\, \bot }}{\left( \frac{\omega_{p}}{c} \right)^3}{\left( \frac{\omega_{p}}{\omega } \right)^2}{\mathscr{F}_{ \bot \left( \parallel  \right)}}\left( \varrho _{\rm{eff}}^{\left( 1 \right)} \right).
\end{equation}

Further, it is convenient to express depolarization factors in terms of  $\varrho _{{\rm{eff}}}^{\left( i \right)}$. Then, the first formula of~(\ref{q3}) takes the form:

\begin{equation}
\label{q19}
\mathcal{L}_\parallel ^{\left( i \right)} = \frac{1}{2}\frac{{\varrho _{{\rm{eff}}}^{\left( i \right)}}^{2}}{{{\left( {1 - \varrho _{{\rm{eff}}}^{\left( i \right)\,2}} \right)}^{{3 \mathord{\left/
 {\vphantom {3 2}} \right.
 \kern-\nulldelimiterspace} 2}}}}\left( \ln \frac{1 + \sqrt {1 - \varrho _{{\rm{eff}}}^{\left( i \right)\,2}} }{1 - \sqrt {1 - \varrho _{{\rm{eff}}}^{\left( i \right)\,2}} } - 2\sqrt {1 - \varrho _{\rm{eff}}^{\left( i \right)\,2}}  \right).
 \end{equation}

It should be pointed out that the formulae for metal-dielectric sphere, given in \cite{b38}, can be obtained from the relations (\ref{q17})--(\ref{q19}) by the limit transition ${e_p} \to 0$. Since $\mathop {\lim }\limits_{\varrho _{{\rm{eff}}}^{\left( 1 \right)} \to 1} \mathcal{L}_{ \bot \left( \parallel  \right)}^{\left( i \right)} = \frac{1}{3}$, and $\mathop {\lim }\limits_{\varrho _{{\rm{eff}}}^{\left( 1 \right)} \to 1} {\mathscr{F}_{ \bot \left( \parallel  \right)}}\left( {\varrho _{{\rm{eff}}}^{\left( 1 \right)}} \right) = \frac{8}{3}$, we obtain:

\begin{equation}
\gamma _{\rm{s}} = \mathscr{A}\left( \omega ,\,\,R \right)\frac{{v_{\rm{F}}}}{R},
\end{equation}

\begin{equation}
\gamma _{{\rm{rad}}}^{ \bot \left( \parallel  \right)} = \frac{1}{{24\piup }}\frac{{{V_0}}}{{\sqrt {{\epsilon_{\rm{m}}}\left( {{\epsilon^\infty } + 2{\epsilon_{\rm{m}}}} \right)} }}{\left( \frac{{{\omega _p}}}{c} \right)^3}{\left( {\frac{{{\omega _p}}}{\omega}} \right)^2}\frac{{{v_{\rm{F}}}}}{R},
\end{equation}
where $V_0$ is the volume of metallic core;

\begin{equation}
\mathscr{A}\left( {\omega ,\,\,R} \right) = \frac{1}{4}{\left( {\frac{{{\omega _p}}}{\omega }} \right)^2},
\end{equation}
${\nu _s} = {{v_{\rm{F}}} \mathord{\left/
 {\vphantom {{{v_{\rm{F}}}} R}} \right.
 \kern-\nulldelimiterspace} R}$, and ${v_{\rm{F}}}$ is the Fermi electron velocity.

In the opposite case ${e_p} \to 1$ \Big($\mathcal{L}_ \bot ^{( i )} = \frac{1}{2}$, $\mathcal{L}_\parallel ^{( i )} = 0$, and ${\mathscr{F}_ \bot }( {\varrho _{{\rm{eff}}}^{( 1 t)}} ) = \frac{{3\piup}}{4}$\Big), the formulae for metal-dielectric cylinder of the infinite length \cite{b34,b38} follow from (2.17)--(2.19):

\begin{equation}
\gamma _{\rm{s}}^\parallel  = \gamma _{{\rm{rad}}}^\parallel  = 0,
\end{equation}

\begin{equation}
\gamma _{\rm{s}}^ \bot  = \frac{{27\piup }}{{128\left( {{\epsilon_{\rm{m}}} + 1} \right)}}{\left( {\frac{{{\omega _p}}}{\omega }} \right)^2}\frac{{{v_{\rm{F}}}}}{a},
\end{equation}

\begin{equation}
\gamma _{\rm{rad}}^ \bot  = \frac{3}{{128\piup }}\frac{V_{0}}{{\sqrt {{\epsilon_{\rm{m}}}\left( {{\epsilon^\infty } + {\epsilon_{\rm{m}}}} \right)} }}{\left( {\frac{{{\omega _p}}}{c}} \right)^3}{\left( {\frac{{{\omega _p}}}{\omega }} \right)^2}\frac{{v_{\rm{F}}}}{a},
\end{equation}
$a$ is the radius of the base of the composite particle core, $V_0$ is the volume of the core.

The absorption cross-section and scattering cross-section are determined by the relations

\begin{equation}
\label{q20}
C_@^{{\rm{abs}}} = \frac{{\omega \sqrt {{\epsilon_{\rm{m}}}} }}{c}\Im \left( {\frac{2}{3}\alpha _@^ \bot  + \frac{1}{3}\alpha _@^\parallel } \right);
 \end{equation}

\begin{equation}
\label{q21}
C_@^{{\rm{sca}}} = \frac{{{\omega ^4}\epsilon_{\rm{m}}^2}}{{6\piup {c^4}}}\left( {\frac{2}{3}{{\left| {\alpha _@^ \bot } \right|}^2} + \frac{1}{3}{{\left| {\alpha _@^\parallel } \right|}^2}} \right)
 \end{equation}
and the radiation efficiency

\begin{equation}
\label{q22}
\xi _@^{{\rm{rad}}} = \frac{C_@^{\rm{abs}}}{C_@^{\rm{abs}} + C_@^{\rm{sca}}}.
 \end{equation}

Subsequently the formulae (\ref{q1}), (\ref{q20})--(\ref{q22}) taking into account the relations (\ref{q2})--(\ref{q4}), (\ref{q12}), (\ref{q13}), (\ref{q15})--(\ref{q19}) are going to be used for  obtaining  the numerical results.

\section{Calculation results and the discussion}

The calculations have been performed for two-layer axisymmetrical nanostructures (prolate spheroids, cylinders and spherocylinders) of  different sizes situated in teflon ($\epsilon_{\rm{m}} = 2.3$). The parameters of metals of the core and shell are given in tables~\ref{tbl-smp1} and~\ref{tbl-smp2}.
%
%
\begin{table}[!t]
	\caption{Parameters of metals (see, for example \cite{b31,b37} and references therein).}
	\label{tbl-smp1}
	\vspace{0.5ex}
	\begin{center}
		\renewcommand{\arraystretch}{0}
		\begin{tabular}{|c|c|c|c|c|c|}
			\hline
			\raisebox{-1.7ex}[0pt][0pt]{Parameters}
			& \multicolumn{5}{c|}{Metals}\strut\\
			\cline{2-6}
			& Cu& Au&  Ag&  Pt&  Pd \strut\\
			\hline
			${r_s}/{a_0}$ &2.11&3.01&3.02&3.27&4.00\strut\\
			\hline
			${m^*}/{m_e}$ &1.49&0.99&0.96&0.54&0.37\strut\\
			\hline
			${\epsilon^\infty }$ &12.03&9.84&3.70&4.42&2.52\strut\\
			\hline
			${\gamma _{{\rm{bulk}}}},\,\,{10^{14}}\,\,{{\rm{s}}^{ - 1}}$\rule{0pt}{2.5ex}
			&0.37&0.35&0.25&1.05&1.39\strut\\
			\hline
		\end{tabular}
		\renewcommand{\arraystretch}{1}
	\end{center}
\end{table}

\begin{table}[!t]
\caption{Dielectric permittivity of the materials of the shell \cite{b39}.}
\label{tbl-smp2}
\vspace{0.5ex}
\begin{center}
\renewcommand{\arraystretch}{0}
\begin{tabular}{|c|c|c|c|c|c|}
\hline
Shell&${\rm{Si}}{{\rm{O}}_2}$&${\rm{A}}{{\rm{l}}_2}{{\rm{O}}_3}$&${\rm{T}}{{\rm{a}}_2}{{\rm{O}}_5}$&
${\rm{Ti}}{{\rm{O}}_2}$&${\rm{Si}}$\strut\\
\hline
${\epsilon_{\rm{s}}}$&2.10&3.10&4.67&6.50&13.69\strut\\
\hline
\end{tabular}
\renewcommand{\arraystretch}{1}
\end{center}
\end{table}

Figure~\ref{fig-smp4} shows the frequency dependencies for the imaginary parts of the longitudinal and transverse components of the polarizability tensor for rod-like nanoparticles $\rm{Au}@\rm{SiO}_2$ of the different shapes. The maxima $\Im \alpha _@^{ \bot \left( \parallel  \right)}\left( \omega  \right)$ correspond to the frequencies of the transverse (longitudinal) surface plasmonic resonances $\omega_{sp}^{\bot\left(\parallel\right)}$. The results of the calculations indicate that $\omega_{sp}^ \bot  > \omega _{sp}^\parallel$ for the nanoparticles of all considered shapes, and moreover, $\Delta\omega _{sp} = \omega _{sp}^ \bot  - \omega _{sp}^\parallel$ increases in the sequence of the shapes ``cylinder $\rightarrow$ spherocylinder $\rightarrow$ prolate spheroid''. Let us point out that $\max \{ {\Im \alpha _@^\parallel }\} > \max \left\{ {\Im \alpha_@^ \bot } \right\}$ for the nanoparticles of all shapes which are under  consideration.

The comparison $\Im \alpha_@^ \bot \left( \omega  \right)$ and $\Im \alpha _@^\parallel \left( \omega  \right)$ for the same particles of the considered shapes is given in figure~\ref{fig-smp5}. The ``blue'' shift of the maxima $\Im \alpha_@^\parallel $ and the ``red'' shift of the maxima $\Im \alpha_@^ \bot $ take place in the sequence of the shapes ``prolate spheroid $\rightarrow$ spherocylinder $\rightarrow$ cylinder'', although these shifts are rather insignificant.

Figures~\ref{fig-smp6} and~\ref{fig-smp7} show the results of the calculations of the frequency dependencies for the absorption cross-section and scattering cross-section for the rod-like nanoparticles $\rm{Au}@\rm{SiO}_2$. It should be pointed out that two maxima of the absorption cross-section and scattering cross-section, which correspond to the longitudinal and transverse surface plasmonic resonances, are significant only for the case of the particle of spheroidal shape with the different major and minor semi-axes. For the case of cylinders and spherocylinders, the splitting of the maxima becomes noticeable for the nanoparticles whose longitudinal and transverse sizes are significantly different. This fact confirms the close location of the frequencies $\omega_{sp}^ \bot $ and $\omega_{sp}^\parallel$ for cylinders (see figure~\ref{fig-smp4}). The reason for such a splitting of SPR frequencies is a decrease of the aspect ratio at the same volume in the sequence of forms ``cylinder $\rightarrow$ spherocylinder $\rightarrow$ prolate spheroid''.

Figures~\ref{fig-smp8} and \ref{fig-smp9} show the frequency dependencies for the absorption cross-section for the rod-like nanoparticles with the cores and shells of different materials. The results of the calculations show that the variation of metal of the core has a much stronger effect both on the splitting of the maxima $C_@^{\rm{abs}}$ and on their location in the spectrum for the nanoparticles of all shapes which are under  consideration. In turn, the variation of the material of the dielectric shell results in only a slight shift of the maxima of the absorption cross-section. This is due to the fact that the frequencies of the longitudinal (transverse) SPR are significantly dependent on the plasma frequency (which is significantly different for different metals) and is practically independent of the dielectric permittivity of the shell.

Figure~\ref{fig-smp10} shows the frequency dependencies for the radiation efficiency for the nanoparticles $\rm{Ag}@\rm{SiO}_2$ of different shapes and sizes. Let us point out that $\xi _@^{{\rm{rad}}}$ is significantly dependent on the sizes of the nanoparticle and is practically independent of its shape. Thus, for the nanoparticles of all shapes, the curves $\xi _@^{{\rm{rad}}}\left( {\hbar \omega } \right)$ are qualitatively similar and quantitatively close, and the radiation efficiency is maximum in the visible (optical) part of the spectrum for the particles with the sizes ${a_l} = 100\,\,{\rm{nm}}$, ${b_t} = 50\,\,{\rm{nm}}$ (when the difference in these sizes is minimum).

\section{Conclusions}

The relations for polarizability and for the absorption and scattering cross-sections of the rod-like metal-dielectric nanoparticles (cylinders of the finite length and spherocylinders) have been obtained in the frameworks of equivalent elongated spheroid approach.

It is shown that the frequencies of transverse SPR are greater than the frequencies of longitudinal SPR for the particles of all considered shapes, and the splitting of the frequencies of SPR increases in the sequence ``cylinder $\rightarrow$ spherocylinder $\rightarrow$ prolate spheroid'', hence, the aspect ratio decreases exactly in this order under the same volume of nanoparticles. Moreover, the maxima of the imaginary parts of the longitudinal component of polarizability tensor are greater than the maxima of the imaginary parts of the transverse component.

It has been established that the frequency dependencies for the light absorption and scattering cross-sections have two maximums which correspond to longitudinal and transverse SPR. The splitting of resonances is noticeable for the particles of spheroidal shape of all sizes, while for the case of cylinders and spherocylinders, the splitting is noticeable only when the longitudinal sizes of the nanoparticle are much greater than its transverse sizes.

It is shown that the variation of the shell material has a weak effect on the location of the maxima of the absorption cross-section, while the variation of metal of the core, for the particles of all considered shapes, has a significant effect both on the value of splitting of resonances and on the location of the maxima on the frequency scale, which is associated with the significant differences in such optical parameters as plasma frequency and the contribution of the interzone transitions into the dielectric function.

The calculations of the radiation efficiency show its strong dependence on the size of the particle and the independence from the shape, while the radiation efficiency is maximum for the particles of all considered shapes, which have the smallest difference in longitudinal and transverse sizes.

\ukrainianpart

\title{Плазмонні ефекти у стрижнеподібних метал-діелектричних наночастинках}
\author{Я.~В.~Карандась\refaddr{label1,label2} }
\addresses{
\addr{label1} Національний університет ``Запорізька політехніка'', вул. Жуковського, 64,
69063 Запоріжжя, Україна
\addr{label2} Запорізький науково-дослідний експертно-криміналістичний центр МВС України, вул. Аваліані 19-А, 69068 Запоріжжя, Україна
}
%
%
%

\makeukrtitle

\begin{abstract}
\tolerance=3000%
З використанням уявлень про еквівалентний витягнутий сфероїд досліджено оптичні
властивості стрижнеподібних двошарових наночастинок. Наведено розрахунки
частотних залежностей поляризованостi та перерізу поглинання і розсiювання
витягнутих сфероїдів, циліндрів та сфероциліндрів. Проаналізовано вплив розмірів
та форм наночастинок, а також матеріалів ядра та оболонки на положення максимумів
уявної частини поляризованості та перерізів поглинання та розсіювання.
Сформульовано рекомендації щодо форми та відношення розмірів наночастинок
із максимальним значенням радіаційної ефективності.
\keywords поляризованiсть, еквівалентний витягнутий сфероїд, частотні залежності, перерізи поглинання і розсіювання, стрижнеподібні наночастинки
\end{abstract}

\end{document}